\input{aipcheck}

\documentclass[
    ,final            
  ]
  {aipproc}

\layoutstyle{8x11double}

\begin{document}

\title{Observable Signatures of the Accretion-Induced Collapse of White Dwarfs}


\author{Brian D. Metzger}{
  address={Astronomy Department and Theoretical Astrophysics Center, University of California, Berkeley, 601 Campbell Hall, Berkeley CA, 94720}
}

\author{Anthony L. Piro}{
  address={Astronomy Department and Theoretical Astrophysics Center, University of California, Berkeley, 601 Campbell Hall, Berkeley CA, 94720}
}

\author{Eliot Quataert}{
  address={Astronomy Department and Theoretical Astrophysics Center, University of California, Berkeley, 601 Campbell Hall, Berkeley CA, 94720}
}

\author{Todd A. Thompson}{
  address = {Department of Astronomy and Center for Cosmology $\&$ Astro-Particle Physics, The Ohio State University, Columbus, OH 43210}
}
\begin{abstract}

Despite its probable occurrence in Nature, the accretion-induced collapse (AIC) of a white dwarf (WD) has not yet been directly observed (or, at least, it has not been unambiguously identified as such).  In this contribution we summarize the observational signatures of AIC, emphasizing its possible role as both an optical and high-energy transient.  Due to the low quantity of ejected mass and radioactive Ni, the isolated collapse of a slowly-rotating WD is difficult to detect much beyond our own Galaxy.  If, however, the progenitor WD is rapidly-rotating (a justifiable assumption), a significant portion of its mass must be shed during collapse into an accretion disk around the newly-formed proto-neutron star (NS).  As a result, the observational manifestations of AIC {\it with rotation} expand considerably, possibly to include: (1) a day-long optical transient with a peak brightness of up to $M_{\rm V} \sim -15$ and a broad-lined, Fe-group-rich spectrum (termed ``naked'' AIC); (2) a $\sim $week-long sub-luminous Type I SN, perhaps similar to SNe 2005e and 2008ha (termed ``enshrouded'' AIC); (3) a strong gravitational wave source; and, possibly, (4) a short-duration gamma-ray burst (or some analogous high-energy transient lasting $\sim 1$ second).  AIC with rotation also produces a rapidly-spinning NS, which has long been considered a promising channel for producing  millisecond pulsars.  If this rapidly-spinning NS forms strongly magnetized (a ``magnetar''), its spin-down luminosity may power a $\sim 100$ second ``flash'' of high-energy radiation via internal shocks in the NS's neutrino-heated wind.  Isolated magnetar birth through AIC provides a natural explanation for the X-ray tails observed following some short GRBs and, potentially viewable over a larger solid angle than the gamma-ray precursor, may also be detectable with proposed all-sky X-ray survey missions such as $EXIST$.  

\end{abstract}

\maketitle

\vspace{-0.4cm}
\section{Introduction}
\vspace{-0.3cm}
As an accreting white dwarf (WD) approaches the Chandrasekhar mass ($M_{\rm ch}$), nuclear fuel ignites in its core.  In many cases this is thought to lead to a thermonuclear explosion and a Type Ia supernova (SN) (e.g. \cite{2000ARA&A..38..191H}).  In some circumstances, however, electron captures may efficiently rob the core of its degeneracy pressure-support, causing core-collapse before an explosion can ensue.  Such an accretion-induced collapse (AIC; \cite{1976A&A....46..229C}) may result both from accretion following Roche-lobe overflow from a non-degenerate companion (the ``single degenerate'' AIC channel; e.g. \cite{1980PASJ...32..303M}), as well as following the merger of two WDs in a binary system \cite{1989A&A...209..111M}.

Despite being a clear-cut theoretical prediction, AIC has not yet been directly, unambiguously observed.  One reason for its absence among the transient census may be that AIC is less common than the ``normal'' evolutionary channels responsible for Type Ia SNe.  Theory suggests, however, that this may not be the whole explanation: AIC is currently considered the most likely outcome of super-$M_{\rm ch}$ WD-WD mergers (e.g. \cite{2007MNRAS.380..933Y}), events which population synthesis calculations find occur at a Galactic rate of $\sim 10^{-3}$ yr$^{-1}$, comparable to the estimated Type Ia SN rate \cite{2009arXiv0904.3108R}.  Such a high AIC rate may, however, be in conflict with their predicted yields of rare neutron-rich isotopes (e.g. \cite{1999ApJ...516..892F}).

\begin{figure}
  \includegraphics[height=.3\textheight]{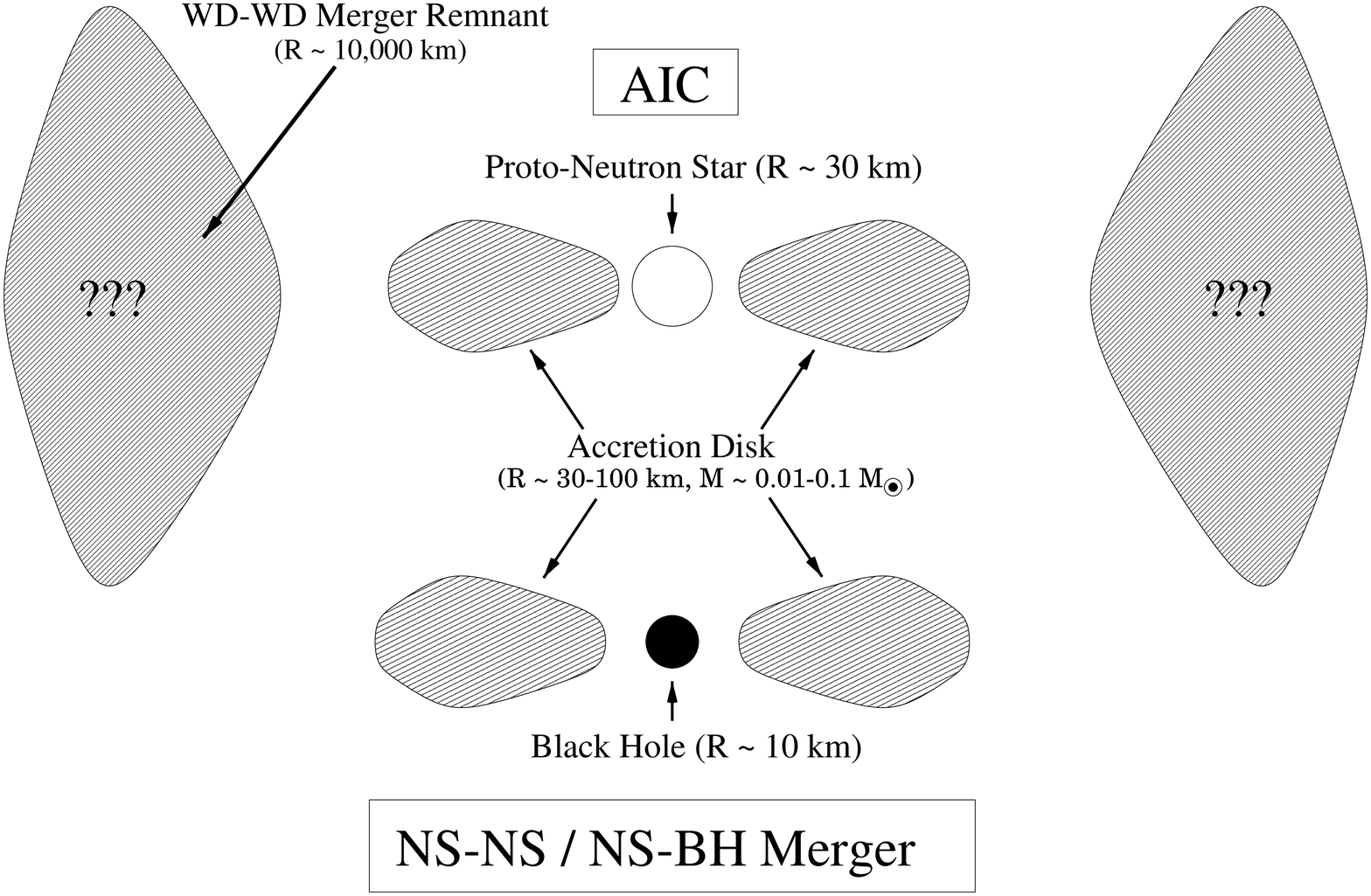}
  \caption{Similar Systems$-$Distinct Origins.  ($Top$) Configuration immediately following the accretion-induced collapse of a WD and ($Bottom$) following the merger of a NS-NS or NS-BH binary.  Despite rather distinct origins, both systems result in a central compact object (NS or BH) surrounded by an accretion disk of characteristic mass $\sim 10^{-2}-0.1$M$_{\odot}$ and size $\sim 30-100$ km.  If AIC occurs following a WD-WD merger, additional centrifugally-supported material may remain at the merger radius ($\sim 10,000$ km).  Through its interaction with outflows from the NS and accretion disk, this remnant torus may affect the duration and spectral characteristics of the optical transients from AIC (see Fig. \ref{fig:lightcurve}).  If the remnant disks remains bound, it will accrete onto the NS on a timescale of $\sim 100$ s, thus providing one explanation for late-time X-ray activity in AIC models for short-duration GRBs.}
\label{fig:fig1}
\end{figure}

AIC may also be undetected thus far because it produces an optical transient that is dimmer and/or evolves faster than normal SNe \cite{1992ApJ...391..228W}.  State-of-the-art core-collapse simulations show that while AIC {\it does} produce a successful SN by the neutrino mechanism (\cite{2006A&A...450..345K}, \cite{2006ApJ...644.1063D}), the quantity of shock-synthesized Ni and the total ejected mass from the proto-NS are tiny: $M_{\rm Ni} \sim 3\times 10^{-4}M_{\odot}$ and $M_{\rm tot} \sim 2\times 10^{-3}M_{\odot}$, respectfully \cite{2006ApJ...644.1063D}.  Due to the resulting low radioactive decay power and the short timescale before gamma-rays escape the expanding ejecta, the ``supernovae'' produced by isolated NS birth are difficult to detect much beyond our own Galaxy.

Crucial, however, is the role that rotation may play in AIC: a WD that reaches $M_{\rm ch}$ must accrete angular momentum as well as mass.  In single degenerate scenarios, the WD rotation period at collapse is probably set by the spin-equilibrium determined by the pre-collapse accretion rate and surface dipole magnetic field strength \cite{1991PhR...203....1B}.  By contrast, for AIC triggered by a WD-WD merger, rapid rotation at collapse is likely inevitable (e.g. \cite{2007MNRAS.380..933Y}).  As we now discuss, rotation qualitatively changes both the strength and variety of AIC's observable manifestations, thus improving its detection prospects considerably.
\vspace{-0.4cm}
\section{Hyper-Accreting Disks: Evolution and Outflows}
\vspace{-0.3cm}
When a rapidly-spinning WD undergoes collapse, the result is a rapidly-spinning neutron star (NS) encircled by a centrifugally-supported disk (e.g. \cite{1987Natur.329..310M},\cite{1990ApJ...353..159B}).  Using 2D axisymmetric calculations, ref.~\cite{2006ApJ...644.1063D},\cite{2007ApJ...669..585D} find that AIC produces a Keplerian disk with size $\sim 30-100$ km and mass $\sim 0.1-0.5 M_{\odot}$, depending on the rotation profile of the progenitor WD.  A cartoon of the basic configuration following AIC is shown in the top of Figure 2.  This illustration highlights the similarities between AIC and another situation of astrophysical interest: the merger of a NS-NS or NS-black hole(BH) binary, which differs essentially from AIC only in that the central object is a BH rather than a NS.    

The disk formed via AIC is subject to MHD turbulence, which causes it to accrete onto the NS.  This occurs on the viscous timescale, given by
\begin{equation}
t_{\rm visc} \approx 1{\,\rm s}\left(\frac{\alpha}{0.1}\right)^{-1}\left(\frac{R}{100{\,\rm km}}\right)^{3/2}\left(\frac{H/R}{0.2}\right)^{-2},
\label{eq:tvisc}
\end{equation}
where $\alpha$, $R$, and $H$ are the disk's viscosity, radius, and scale-height, respectively.  The large masses and short viscous times associated with the disks produced by AIC and NS-NS/NS-BH mergers imply very large accretion rates ($\sim M_{\odot}$s$^{-1}$).  Such disks are termed ``hyper-accreting'' and have been extensively studied, particularly in the context of gamma-ray burst (GRB) progenitors (e.g. see \cite{2008AIPC.1054...51B} and \cite{lr07}, and references therein).

Formed with a temperature of several MeV, the disk's composition is initially broken down into free nuclei, with the neutron-to-proton ratio $n/p$ set by an equilibrium between the pair capture reactions (e.g. \cite{2003ApJ...586.1254P})  
\begin{equation}
e^{-}+p \rightarrow \nu_{e}+n; {\,\,\,\,}e^{+}+n \rightarrow \bar{\nu_{e}}+p,
\label{eq:paircaptures}
\end{equation}
which are rapid at high temperatures.  Since the disk is initially neutrino cooled and electron degenerate, $n/p$ is regulated to a value $\sim 10$ \cite{2007ApJ...657..383C}.  

\begin{table}
\begin{tabular}{lllll}
\hline
  \tablehead{1}{l}{b}{Astrophysical Source} &
  \tablehead{1}{l}{b}{Entropy} & 
  \tablehead{1}{l}{b}{Expansion Timescale} &
  \tablehead{1}{l}{b}{Electron Fraction\tablenote{Electron fraction is defined as $Y_{e} \equiv p/(n+p)$, where $n/p$ are the neutron/proton densities.}} &
  \tablehead{1}{l}{b}{Product Nuclei} \\
 & ($k_{\rm b}$ baryon$^{-1}$) & (s) & & \\
\hline
Big Bang Nucleosynthesis & $\sim 10^{10}$ & $\sim 10^{2}$ & 0.88 & H, He, D, Li \\
Gamma-Ray Burst Fireball\tablenote{\cite{2002ApJ...580..368P}} & $\sim 10^{5}$ & $10^{-3}-10^{-2}$ & $0.1-0.6$ & H, n, D \\ 
Neutrino-Heated Winds\tablenote{Originating both from proto-neutron stars following core-collapse SNe \cite{1996ApJ...471..331Q} and from hyper-accreting disks \cite{2006ApJ...643.1057S},\cite{2008ApJ...676.1130M}.} & $\sim 10^{2}$ & $10^{-3}-0.1$ & $0.4-0.6$ & $\alpha$-process; $r$-process (?)\\
Hyper-Accreting Disk Disruption:& & & & \\
{\,\,\,}{\it NS-NS/NS-BH Mergers\tablenote{\cite{2008MNRAS.390..781M},\cite{2009MNRAS.396..304M}}} & $\sim 10$ & $\sim 1$ & $0.1-0.4$ & n-rich NSE; $r$-process\\
{\,\,\,}{\it Accretion-Induced Collapse}\tablenote{\cite{2009MNRAS.tmp..653M}} & {$\sim 10$} & {$\sim 1$} & {$0.4-0.6$} & {$^{56}$Ni; n-rich NSE} \\
Dynamical Ejecta from & $\le 1$ & $10^{-3}-10^{-2}$ & $\sim 0.1$ & $r$-process \\
{\,\,\,}NS-NS/NS-BH Mergers\tablenote{e.g. \cite{1999ApJ...525L.121F}} & & & & \\
\hline
\end{tabular}
\caption{Astrophysical Sites of Nucleosynthesis which Undergo Freeze-Out from Nuclear Statistical Equilibrium (NSE)}
\label{tab:a}
\end{table}

As the disk accretes, the bulk of its remaining mass $M_{\rm D}$ spreads to larger radii in order to conserve angular momentum $\propto M_{\rm D}R^{1/2}$.  After a few seconds of viscous evolution, the disk's outer edge expands to $\sim 10^{3}$ km as its mass decreases to $\sim 1/3$ of its initial value.  At this point, when the temperature drops below $\sim 1$ MeV, several important changes occur nearly simultaneously (\cite{2009MNRAS.396..304M}; see also \cite{2008AIPC.1054...51B}): 
\begin{enumerate}
\item{The disk midplane transitions from being gas pressure-supported and mildly electron-degenerate to being radiation pressure-supported and non-degenerate.}
\item{Neutrino cooling becomes ineffective at offsetting viscous heating; the disk thickens and becomes only marginally bound to the central object.}
\item{Weak interactions in the disk (e.g. eq.~[\ref{eq:paircaptures}]) "freeze out," i.e. $n/p$ stops evolving.}
\item{Free nuclei recombine into alpha-particles, since the latter are favored in nuclear statistical equilibrium (NSE) at low temperatures.}
\end{enumerate}

The energy released when alpha-particles form ($\sim 7$ MeV nucleon$^{-1}$) is more than sufficient to unbind the disk given its large radius and marginally-bound state.  {\bf Thus, an inevitable consequence of the viscous evolution of a massive, hyper-accreting disk (Fig.~1) is the ultimate re-ejection of a significant fraction ($\sim 20-50\%$) of the disk's mass back into space} (\cite{2008MNRAS.390..781M},\cite{2009MNRAS.396..304M},\cite{2008AIPC.1054...51B},\cite{2009MNRAS.392.1451R}).  This conclusion was recently confirmed by multi-dimensional numerical simulations \cite{2009arXiv0904.3752L}.

Since the resulting ejecta is hot and dense (entropy $\sim 10$ $k_{\rm b}$ baryon$^{-1}$) with a relatively slow expansion timescale $\sim 0.3-3$ s (set by the disk's viscous time as the outflow begins), the outflow will efficiently recombine into heavy nuclei as it quasi-adiabatically expands \cite{1992ApJ...395..202W}.  This represents a previously-unexplored site of heavy element nucleosynthesis \cite{2009MNRAS.396..304M}.

Table \ref{tab:a} presents a comparison of the outflow properties and nucleosynthetic yields of various astrophysical sites that undergo a freeze-out from NSE.  This comparison highlights the unique parameter space occupied by the outflows from ``hyper-accreting disk disruption'' in relation to other sites of heavy element nucleosynthesis, such as its position in entropy between the ``cold'' NS material dynamically-ejected during NS-NS/NS-BH mergers \cite{1999ApJ...525L.121F} and the ``high-entropy'' neutrino-heated winds from proto-neutron stars \cite{1996ApJ...471..331Q} and earlier stages in the evolution of hyper-accreting disks \cite{2006ApJ...643.1057S},\cite{2008ApJ...676.1130M}.  

\begin{figure}
  \includegraphics[height=.3\textheight]{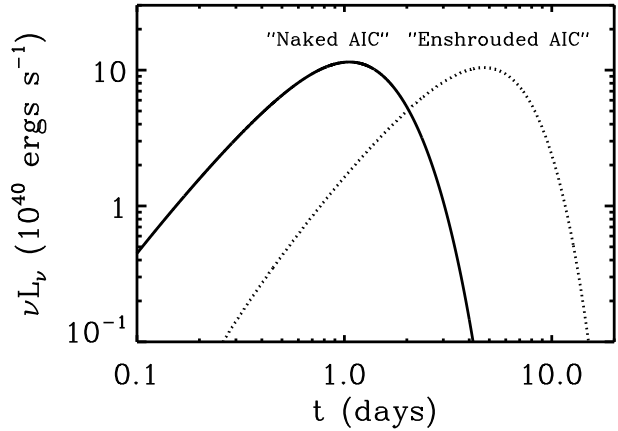}
  \caption{Example V-band Light Curves of the Optical Transients from AIC \cite{2009MNRAS.tmp..653M}.  The solid line shows an example of a ``Naked AIC'' event with Ni mass $M_{\rm Ni} = 10^{-2}M_{\odot}$, velocity $v = 0.1$ c, and total ejected mass $M_{\rm tot} = 2\times 10^{-2}M_{\odot}$ (as expected if the ejected disk material provides all of the outflow's opacity).  The dashed line shows an ``Enshrouded AIC'' event with the same Ni mass ($ = 10^{-2}M_{\odot}$) but with a larger total mass ($= 0.2M_{\odot}$) and slower velocity ($v = 10,000$ km s$^{-1}$), as expected if the Nickel wind efficiently shares its mass and energy with the remnant torus from a WD-WD merger (see Fig.~\ref{fig:fig1}).  Enshrouded AIC is a possible explanation for sub-luminous Type I SNe such as SN 2008ha (\cite{2009Natur.459..674V},\cite{2009arXiv0902.2794F}) and 2005e (\cite{2009arXiv0906.2003P}).  Both naked and enshrouded AIC are targets for upcoming rapid-cadence optical transient surveys.
}
\label{fig:lightcurve}
\end{figure}
\vspace{-0.4cm}
\subsection{Nucleosynthetic Yields}
\vspace{-0.3cm}
In the case of black hole accretion (as applicable to NS-NS/NS-BH mergers) the disk freezes out moderately neutron-rich with $n/p \sim 2-3$ \cite{2008MNRAS.390..781M},\cite{2009MNRAS.396..304M}.  This occurs because (1) pair captures (eq.~[\ref{eq:paircaptures}]) remain the dominant weak reaction through freeze-out and (2) the disk's degeneracy is lifted only {\it as freeze-out is occurring}.  As a result, late-time outflows from NS-NS/NS-BH merger disks synthesize heavy ($A \ge 70$) neutron-rich isotopes (e.g. $r$-process elements; \cite{2009MNRAS.396..304M}).  Since the amount of ejected material per event is substantial ($\sim 10^{-3}-10^{-2}M_{\odot}$) and these elements are very rare, merging compact objects may have an important effect on the chemical evolution of the Galaxy.  Indeed, estimates of the amount of neutron-rich material ejected per event can be used to place meaningful constraints on the Galactic NS-NS/NS-BH merger rate \cite{2009MNRAS.396..304M}.

A crucial difference in the case of AIC is the presence of the central proto-NS, which irradiates the disk with a powerful flux of electron neutrinos as the NS deleptonizes.  By calculating the viscous evolution of AIC disks, we find that $\nu_{e}$-irradiation raises $n/p$ at freeze-out to $\sim 1$ because neutrino absorptions ($\nu_{e}+n\rightarrow e^{-}+p$) overcome the electron captures which otherwise cause a neutron-rich freeze-out \cite{2009MNRAS.tmp..653M}.  Freeze-out with $n/p \sim 1$ is crucial to detecting AIC because dense proton-rich ejecta is efficiently synthesized into radioactive $^{56}$Ni \cite{2008ApJ...685L.129S}.
\vspace{-0.3cm}
\section{Optical Transients from AIC}
\vspace{-0.3cm}
\subsection{"Naked" AIC}
\vspace{-0.3cm}
If a disk with mass $\sim 0.1 M_{\odot}$ forms during AIC, our calculations predict that the total ejected mass is $M_{\rm tot} \sim 3\times 10^{-2}M_{\odot}$, of which $M_{\rm Ni} \sim 10^{-2}M_{\odot}$ produces $^{56}$Ni.  More generally, the disk mass (and hence the ejected mass) depends on the rotation rate of the WD at collapse, which could vary substantially between events.  The speed of this ``Ni wind'' $v_{\rm Ni} \sim 0.1$ c is set by the $\sim 8$ MeV nucleon$^{-1}$ released in forming heavy elements.  As in Type Ia SNe, the Nickel's decay will reheat the (adiabatically cooled) ejecta sufficiently to produce detectable transient emission once the outflow expands sufficiently that photons can diffuse out \cite{2009MNRAS.tmp..653M}.  Figure \ref{fig:lightcurve} shows a calculation of the expected V-band light curve, which in this case peaks on a timescale $\sim 1$ day at a luminosity of $\sim 10^{41}$ ergs s$^{-1}$ ($M_{V} \sim -14$).  

We refer to these events as ``naked AIC'' transients because the mass ejected from the disk wind alone is relatively small, and it is assumed to interact with little additional material as it expands to large radii.  This situation may best describe the environments characterizing single degenerate AIC scenarios.  Although significantly fainter and faster-evolving than normal SNe, naked AIC is a promising target for upcoming rapid-cadence optical transient surveys such as the PanSTARRs Medium Deep Survey (MDS; \cite{Kaiser}), the Palomar Transient Factory (PTF; \cite{Rau}), and the Large Synoptic Survey Telescope (LSST).  Importantly, as AIC with rotation may also be a strong source of gravitational waves \cite{2005ApJ...625L.119O}, this radioactively-powered transient may provide the ``beacon'' which is crucial to confirming and localizing the gravitational signal \cite{2009astro2010S.235P}.

Since $^{56}$Ni is likely to be the dominant isotope synthesized in AIC disk winds, an Fe group-rich, broad-lined spectrum is one of a naked AIC's defining characteristics.  However, other heavy elements (e.g. Cr, Kr, Se, Br) will also be produced in smaller abundances because the disk's ejecta possesses a distribution of $n/p$ in the range $\sim 0.5-1.5$ (see Fig.~2 of \cite{2009MNRAS.tmp..653M}). A ``smoking gun'' would be the identification of unusual spectral features from these rare elements.  
\begin{figure}
  \includegraphics[height=.3\textheight]{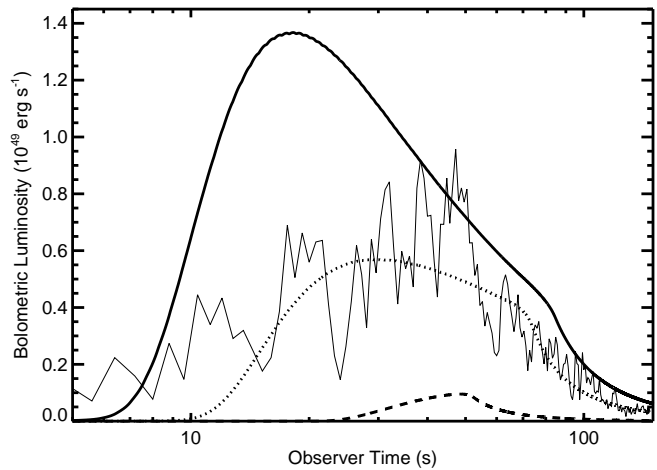}
  \caption{From \cite{2008MNRAS.385.1455M}.  Luminosity of internal shock emission in proto-magnetar winds, calculated for a NS with an initial rotation period $P_{0} = 1$ ms and for three values of the surface dipole field strength (from bottom to top) $B = 10^{15}$ G, $3\times 10^{15}$ G, and $10^{16}$ G.  An electron acceleration efficiency $\epsilon_{e} = 0.5$ is assumed.  Note the lack of emission at early times because the outflow is non-relativistic.  The gradual onset of the emission is due to the large Thomson optical depth, which decreases as the outflow expands.  The late-time decline in emission is the onset of curvature emission from the last shock, produced by the shell released at $t = 40$ s, after which the NS becomes optically-thin to neutrinos.  The late-time {\it Swift} BAT light curve from GRB060614 from \cite{2007ApJ...663..407B}, shown with a light solid line and scaled to the physical isotropic luminosity, is reproduced in a time-averaged sense by the $B = 3\times 10^{15}$ G model.}
\label{fig:spindown}
\end{figure}
\vspace{-0.4cm}
\subsection{"Enshrouded" AIC}
\vspace{-0.3cm}
If AIC results from a WD-WD merger, a significant amount of C, O, Ne, or He ($\sim 0.1M_{\odot}$) may remain near the radius ($\sim 10^{9}$ cm) characterizing the merger.  Once the outgoing SN from the NS or the Ni wind from the accretion disk (eacuh with energy $\sim 10^{50}$ ergs) impacts this remnant torus, it will (1) shock heat the material to a few times $10^{9}$ K, which may synthesize intermediate mass elements, but will likely leave some unburned WD material; (2) slow the ejecta from $v \sim 30,000$ km s$^{-1}$ to $\sim 3-10,000$ km s$^{-1}$, depending on how efficiently the Nickel wind's mass and energy couple to the remnant disk.  A combination of slower ejecta and higher opacity will lengthen the duration of the light curve $\propto M_{\rm tot}^{1/2}v^{-1/2}$ (but without strongly changing the peak brightness), making such an ``enshrouded AIC'' easier to detect than naked AIC.  Figure \ref{fig:lightcurve} compares the light curves produced by naked and enshrouded AIC for a case in which the torus is $\sim 10$ times as massive as the Ni-wind and the coupling efficiency is assumed to be $\sim 1$.

Enshrouded AIC represents a possible explanation for recently-discovered sub-luminous, sub-Chandrasekhar Type I SNe, such as 2005e (\cite{2009arXiv0906.2003P}) 2008ha (\cite{2009Natur.459..674V},\cite{2009arXiv0902.2794F}).  For instance, 2008ha rose to its peak brightness in only $\sim 10$ days, possessed low line velocities ($\sim 2000$ km s$^{-1}$), and was inferred to have $M_{\rm Ni} \approx 3\times 10^{-3}M_{\odot}$ and $M_{\rm tot} \sim 0.15M_{\odot}$ \cite{2009arXiv0902.2794F}.  A model similar to enshrouded AIC was proposed by ref.~\cite{1995ASPC...72..164N} to explain the faint SN 1991b.  We emphasize, however, that without a Ni-rich outflow from the accretion disk, current calculations of the shock-synthesized Ni from AIC explosions (e.g. \cite{2006ApJ...644.1063D}), as invoked by this model, could not explain even the faintest SNe.  

There is significant room for the future development of the enshrouded AIC model.  In particular, multi-dimensional hydrodynamic simulations may be required to determine the efficiency with which the SN shock or Ni wind couples to the remnant torus, as well as to quantitatively address the shocked-torus nucleosynthesis.  Interestingly, since a portion of the fast-moving Ni wind may escape through the polar regions without coupling to the torus, both an (early-time) naked and (late-time) enshrouded AIC transient could be produced by the same event, possibly resulting in a double-peaked light curve.
\vspace{-0.4cm}
\section{High Energy Transients}
\vspace{-0.3cm}
\subsection{Short-Duration Gamma-Ray Bursts}
\vspace{-0.3cm}
\label{sec:SGRBEEs}

Given the ubiquitous association of jets with accretion disks in astrophysics \cite{1999PhR...311..225L}, it is tempting to speculate that NS accretion following AIC may also produce a collimated relativistic outflow.  If so, such an outflow's characteristic duration is probably the viscous timescale $t_{\rm visc} \sim 1$ s (eq.~[\ref{eq:tvisc}]).  Of the known high-energy transients, short-duration gamma-ray bursts (GRBs) provide the closest match to the expected energetics and timescale of AIC accretion.  This is not surprising considering the similarity between AIC and NS-NS/NS-BH mergers (see Fig.~\ref{fig:fig1}), the currently popular model for short GRBs (e.g. \cite{lr07}).  It thus seems reasonable to include AIC as a possible short GRB progenitor.  Such a hypothesis is supported by the discovery that short GRBs originate from both early- and late-type galaxies (e.g. \cite{2006ApJ...638..354B},\cite{Barthelmy05}), qualitatively consistent with the expected population of massive WD binaries \cite{2009arXiv0904.3108R}.

\vspace{-0.6cm}
\subsection{X-Ray Flashes from Magnetar Spin-Down}
\vspace{-0.3cm}
One of the biggest mysteries associated with short GRBs is that $\sim 25\%$ of those detected by {\it Swift} are followed by an X-ray tail of extended emission starting $\sim 10$ s after the GRB and lasting for $\sim 10-100$ s (e.g. \cite{NB06}).  This emission indicates that the central engine is active at very late times ($t \gg t_{\rm visc}$), which is difficult to explain in NS-NS/NS-BH merger models because the disk blows apart at late times (\cite{2008MNRAS.390..781M}; see, however, \cite{2009arXiv0904.3752L}).  

In the AIC-GRB model there are, by contrast, a number of plausible mechanisms for producing late-time emission on this timescale.  If the coupling between the Ni wind and the remnant torus discussed previously is not completely efficient, a portion of the torus may remain bound to the NS.  This would allow it to accrete onto the NS at later times, which could power a bipolar jet similar to that produced during the prompt accretion episode, but after a delay $\sim 10-100$ s reflecting the accretion timescale at the torus radius $\sim 10^{9}$ cm \cite{2008MNRAS.385.1455M}. 

Another possibility for powering extended emission is to tap into the (substantial) rotational energy of the NS.  The rapidly-spinning NSs produced by AIC have long been considered promising channels for producing millisecond pulsars (e.g. \cite{1990ApJ...353..159B}).  If the NS is created with a strong magnetic field (a ``magnetar''; \cite{DT92}), late-time high-energy emission could also be powered by the NS's electromagnetic spin-down \cite{1992Natur.357..472U}.  

We have calculated \cite{2008MNRAS.385.1455M} the time-evolution of the mass- and energy-loss rates ($\dot{M}$ and $\dot{E}$) of the outflows from magnetars during the first $\sim 100$ s after their formation using MHD wind calculations that include the relevant neutrino heating and cooling processes \cite{2007ApJ...659..561M}.  Our calculations show that proto-magnetar winds possess a significant reservoir of ``free energy'' because their Lorentz factor $\propto \dot{E}/\dot{M}$ increases monotonically with time as the NS cools.  This naturally leads to the production of ``internal shocks'' when faster material ejected at late times catch up with the slower material ejected earlier.  Figure \ref{fig:spindown} shows our calculation of the (time-averaged) bolometric luminosity from internal shocks in a proto-magnetar's outflow, calculated for a NS with an initial rotation period $P_{0} = 1$ ms and for three values of the surface magnetic dipole field strength $B$.  Also shown is the late-time {\it Swift} BAT light curve of the extended emission from GRB060614 (from \cite{2007ApJ...663..407B}), which is reasonably reproduced by our model with $B = 3\times 10^{15}$ G. 

Relative to the prompt GRB (which is powered by the jet from the prior accretion phase), the magnetar wind's luminosity may be relatively isotropic, with a large solid angle accessible to observers.  Thus, while viewers along the rotation axis observe a short GRB with an extended X-ray tail, equatorial viewers may observe just an X-ray tail with a weak or no gamma-ray precursor.  Isolated magnetar birth thus provides a promising target for proposed all-sky X-ray survey missions such as EXIST \cite{2009astro2010S.278S}.


\vspace{-0.5cm}
\begin{theacknowledgments}
\vspace{-0.3cm}
  I thank the organizers for a fruitful conference that brought together the neutron star and GRB communities in an exciting location.  In particular, I acknowledge helpful conversations with N.~Bucciantini, M.~Cantiello, J.~Grindlay, J.~Lattimer, A.~Lazar,  A.~MacFadyen, P.~Mazzali, B.~Stevens, C.~Thompson, and A.~van Marle.
\end{theacknowledgments}
\vspace{-0.2cm}


\bibliographystyle{aipproc}   
\vspace{-0.3cm}

\bibliography{sample}

\begin{thebibliography}{18}
\expandafter\ifx\csname natexlab\endcsname\relax\def\natexlab#1{#1}\fi
\providecommand{\enquote}[1]{``#1''}
\expandafter\ifx\csname url\endcsname\relax
  \def\url#1{\texttt{#1}}\fi
\expandafter\ifx\csname urlprefix\endcsname\relax\def\urlprefix{URL }\fi
\providecommand{\eprint}[2][]{\url{#2}}



\bibitem[{Hillebrandt 
\& Niemeyer}(2000)]{2000ARA&A..38..191H} {Hillebrandt}, W., \& {Niemeyer}, J.~C.\ 2000, \emph{ARAA}, 38, 191 

\bibitem[Canal 
\& Schatzman(1976)]{1976A&A....46..229C} Canal, R., \& Schatzman, E.\ 1976, \emph{A$\&$A}, 46, 229 

\bibitem[Miyaji et al.(1980)]{1980PASJ...32..303M} Miyaji, S. et al.\ 1980, \emph{PASJ}, 32, 303 

\bibitem[Mochkovitch \& Livio(1989)]{1989A&A...209..111M} Mochkovitch, R., \& Livio, M.\ 1989, \emph{A$\&$A}, 209, 111 

\bibitem[Yoon et al.(2007)]{2007MNRAS.380..933Y} Yoon, S.-C. et al.\ 2007, \emph{MNRAS}, 380, 933 

\bibitem[Ruiter et al.(2009)]{2009arXiv0904.3108R} Ruiter, A.~J. et al.\ 2009, \eprint{arXiv:astro-ph/0904.3108} 

\bibitem[Fryer et al.(1999)]{1999ApJ...516..892F} Fryer, C., Benz, W., 
Herant, M., \& Colgate, S.~A.\ 1999, \emph{ApJ}, 516, 892 

\bibitem[Woosley \& Baron(1992)]{1992ApJ...391..228W}
Woosley, S. E. \& Baron, E. 1992, \emph{ApJ}, 391, 228

\bibitem[Kitaura et 
al.(2006)]{2006A&A...450..345K} Kitaura, F.~S. et al.\ 2006 \emph{A$\&$A}, 450, 345 

\bibitem[Dessart et al.(2006)]{2006ApJ...644.1063D} Dessart, L.~et al.\ 2006, \emph{ApJ}, 644, 1063 

\bibitem[Dessart et al.(2007)]{2007ApJ...669..585D} Dessart, L. et al.\ 2007, \emph{ApJ}, 669, 585 

\bibitem[Bhattacharya 
\& van den Heuvel(1991)]{1991PhR...203....1B} Bhattacharya, D., \& van den Heuvel, E.~P.~J.\ 1991, \emph{Phys.~Rep}, 203, 1 

\bibitem[Michel(1987)]{1987Natur.329..310M} Michel, F.~C.\ 1987, \emph{Nature}, 329, 310

\bibitem[Bailyn 
\& Grindlay(1990)]{1990ApJ...353..159B} Bailyn, C.~D., \& Grindlay, J.~E.\ 1990, \emph{ApJ}, 353, 159 

\bibitem[Beloborodov(2008)]{2008AIPC.1054...51B} Beloborodov, A.~M.\ 2008, 
AIP Conf.~Series, 1054, 51 

\bibitem[Metzger et al.(2009)]{2009MNRAS.396..304M} Metzger, B.~D., Piro, 
A.~L., \& Quataert, E.\ 2009, \emph{MNRAS}, 396, 304 

\bibitem[Lee \& Ramirez-Ruiz(2007)]{lr07} Lee, W. \& Ramirez-Ruiz, E. 2007, New J. Phys., 9, 17 

\bibitem[Pruet et al.(2002)]{2002ApJ...580..368P} Pruet, J. et al.\ 2002, \emph{ApJ}, 580, 368 

\bibitem[Pruet et al.(2003)]{2003ApJ...586.1254P} Pruet, J. et al.\ 2003, \emph{ApJ}, 586, 1254 

\bibitem[Chen \& Beloborodov(2007)]{2007ApJ...657..383C} Chen, W.-X., \& 
Beloborodov, A.~M.\ 2007, \emph{ApJ}, 657, 383

\bibitem[Freiburghaus et al.(1999)]{1999ApJ...525L.121F} Freiburghaus, C. et al.\ 1999, \emph{ApJL}, 525, L121 

\bibitem[Qian \& Woosley(1996)]{1996ApJ...471..331Q} Qian, Y.-Z., \& 
Woosley, S.~E.\ 1996, \emph{ApJ}, 471, 331

\bibitem[Surman et al.(2006)]{2006ApJ...643.1057S} Surman, R. et al.\ 2006, \emph{ApJ}, 643, 1057 

\bibitem[Metzger et al.(2008b)]{2008ApJ...676.1130M} Metzger, B.~D., 
Thompson, T.~A., \& Quataert, E.\ 2008b, \emph{ApJ}, 676, 1130 

\bibitem[Metzger et al.(2008)]{2008MNRAS.390..781M} Metzger, B.~D., Piro, 
A.~L., \& Quataert, E.\ 2008, \emph{MNRAS}, 390, 781

\bibitem[Metzger et al.(2009)]{2009MNRAS.tmp..653M} Metzger, B.~D. et al.\ 2009, \emph{MNRAS}, 653 

\bibitem[Rossi \& Begelman(2009)]{2009MNRAS.392.1451R} Rossi, E.~M., \& Begelman, M.~C.\ 2009, \emph{MNRAS}, 392, 1451 

\bibitem[Lee et al.(2009)]{2009arXiv0904.3752L} Lee, W.~H. et al.\ 2009, \eprint{arXiv:astro-ph/0904.3752}

\bibitem[Woosley 
\& Hoffman(1992)]{1992ApJ...395..202W} Woosley, S.~E., \& Hoffman, R.~D.\ 1992, \emph{ApJ}, 395, 202 

\bibitem[Seitenzahl et al.(2008)]{2008ApJ...685L.129S} Seitenzahl, I.~R. et al.\ 2008, \emph{ApJL}, 685, L129 

\bibitem[Kaiser et al.(2002)]{Kaiser} Kaiser, N.~et al.~2002 Proc.~of the SPIE, 4836, 154

\bibitem[Rau et al.(2008)]{Rau} Rau, A., et al. 2008, \eprint{http://www.weizmann.ac.il/home/galyam/PTF/Rau-paper.pdf}

\bibitem[Ott et al.(2005)]{2005ApJ...625L.119O} Ott, C.~D. et al.\ 2005, \emph{ApJL}, 625, L119 

\bibitem[Phinney(2009)]{2009astro2010S.235P} Phinney, E.~S.\ 2009, 
Astronomy, 2010, 235 

\bibitem[Metzger et al.(2008)]{2008MNRAS.385.1455M} Metzger, B.~D., 
Quataert, E., \& Thompson, T.~A.\ 2008, \emph{MNRAS}, 385, 1455 

\bibitem[Butler \& Kocevski(2007)]{2007ApJ...663..407B} Butler, N.~R., \& 
Kocevski, D.\ 2007, \emph{ApJ}, 663, 407 

\bibitem[Valenti et al.(2009)]{2009Natur.459..674V} Valenti, S., et al.\ 
2009, \emph{Nature}, 459, 674 

\bibitem[Foley et al.(2009)]{2009arXiv0902.2794F} Foley, R.~J., et al.\ 
2009, \eprint{arXiv:0902.2794}

\bibitem[Perets et al.(2009)]{2009arXiv0906.2003P} Perets, H.~B., et al.\ 
2009, \emph{Nature} submitted, \eprint{arXiv:0906.2003}

\bibitem[Nomoto et al.(1995)]{1995ASPC...72..164N} Nomoto, K. et al.\ 1995, MS Pulsars.~ A Decade of Surprise, 72, 164 

\bibitem[Livio(1999)]{1999PhR...311..225L} Livio, M.\ 1999, \emph{PhysRep}, 311, 225 

\bibitem[{Barthelmy et al.}(2005)]{Barthelmy05}
S.~D. {Barthelmy et al.}, \emph{Nature} \textbf{438}, 994--996 (2005).

\bibitem[Bloom et al.(2006)]{2006ApJ...638..354B} Bloom, J.~S., et al.\ 
2006, \emph{ApJ}, 638, 354 

\bibitem[{Norris} and {Bonnell}(2006)]{NB06}
J.~P. {Norris}, and J.~T. {Bonnell}, \emph{ApJ} \textbf{643}, 266--275 (2006).

\bibitem[{Duncan} and {Thompson}(1992)]{DT92}
R.~C. {Duncan}, and C.~{Thompson}, \emph{ApJ} \textbf{392}, L9--L13 (1992)

\bibitem[Usov(1992)]{1992Natur.357..472U} Usov, V.~V.\ 1992, \emph{Nature}, 357, 472 

\bibitem[Metzger et al.(2007)]{2007ApJ...659..561M} Metzger, B.~D., 
Thompson, T.~A., \& Quataert, E.\ 2007, \emph{ApJ}, 659, 561 

\bibitem[Soderberg(2009)]{2009astro2010S.278S} Soderberg, A.~M.\ 2009, 
Astronomy, 2010, 278 


\end{thebibliography}

\IfFileExists{\jobname.bbl}{}
 {\typeout{}
  \typeout{******************************************}
  \typeout{** Please run "bibtex \jobname" to optain}
  \typeout{** the bibliography and then re-run LaTeX}
  \typeout{** twice to fix the references!}
  \typeout{******************************************}
  \typeout{}
 }

\end{document}